\documentclass[reprint,amsmath,amssymb,aps,prb]{revtex4-1}

\usepackage{graphicx}
\usepackage{dcolumn}
\usepackage{bm}% bold math
\usepackage{hyperref}% add hypertext capabilities
\usepackage{physics}
\usepackage{tabularx}
\usepackage{amsmath}
\usepackage{amssymb} %maths
\usepackage[utf8]{inputenc} 
\usepackage{xcolor}

\newcommand{\kv}{{\ensuremath{\bm{k}}}}

\newcommand{\qv}{{\ensuremath{\bm{q}}}}
\newcommand{\vv}{{\ensuremath{\bm{v}}}}

\newcommand{\one}{{\ensuremath{n\bm{k}}}}
\newcommand{\two}{{\ensuremath{m\bm{k}+\bm{q}}}}
\newcommand{\Pot}{{\ensuremath{S_{{n \bm{k}}}^{m \bm{k} +\bm{q} }}}}

\newcommand{\kperp}{\ensuremath{k_\perp}}
\newcommand{\kpara}{\ensuremath{k_\parallel}}
\newcommand{\mperp}{\ensuremath{m^*_\perp}}
\newcommand{\mpara}{\ensuremath{{m^*_\parallel}}}

\newcommand{\Eg}{\ensuremath{E_g}}

\newcommand{\Dou}{\ensuremath{{\Xi^{\rm opt}_{u}}}}
\newcommand{\Dod}{\ensuremath{{\Xi^{\rm opt}_{d}}}}
\newcommand{\Dau}{\ensuremath{{\Xi^{\rm ac}_{u}}}}
\newcommand{\Dad}{\ensuremath{{\Xi^{\rm ac}_{d}}}}
\newcommand{\kdotp}{{\ensuremath{\bm{k}\cdot\bm{p}}}}
\newcommand{\pdd}[2]{\ensuremath{\frac{\partial {#1}}{\partial {#2}}}}

\begin{document}

\title{Dominant electron-phonon scattering mechanisms in $n$-type PbTe from first principles}

\author{Jiang Cao\textsuperscript{1}}
\email{jiang.cao@tyndall.ie}
\author{Jos\'e D. Querales-Flores\textsuperscript{1}}
\author{Aoife R. Murphy\textsuperscript{1,2}}
\author{Stephen Fahy\textsuperscript{1,2}}
\author{Ivana Savi\'c\textsuperscript{1}}
\email{ivana.savic@tyndall.ie}
\affiliation{
\textsuperscript{\normalfont{1}}Tyndall National Institute, Dyke Parade, Cork T12 R5CP, Ireland}
\affiliation{
\textsuperscript{\normalfont{2}}Department of Physics, University College Cork, College Road, Cork T12 K8AF, Ireland}

\date{\today}

\begin{abstract}

We present an \emph{ab-initio} study that identifies the main electron-phonon scattering channels in $n$-type PbTe. We develop an electronic transport model based on the Boltzmann transport equation within the transport relaxation time approximation, fully parametrized from first-principles calculations that accurately describe the dispersion of the electronic bands near the band gap. Our computed electronic mobility as a function of temperature and carrier concentration is in good agreement with experiments. We show that longitudinal optical phonon scattering dominates electronic transport in $n$-type PbTe, while acoustic phonon scattering is relatively weak. We find that scattering due to soft transverse optical phonons is by far the weakest scattering mechanism, due to the symmetry-forbidden scattering between the conduction band minima and the zone center soft modes. Soft phonons thus play the key role in the high thermoelectric figure of merit of $n$-type PbTe: they do not degrade its electronic transport properties although they strongly suppress the lattice thermal conductivity. Our results suggest that materials like PbTe with soft modes that are weakly coupled with the electronic states relevant for transport may be promising candidates for efficient thermoelectric materials.

\end{abstract}

\maketitle

%%%%%%%%%%%%%%%%%%%%%%%%%%%%%%%%%%%%%%%%%%
%%   Introduction
%%%%%%%%%%%%%%%%%%%%%%%%%%%%%%%%%%%%%%%%%%
\section{Introduction}

Lead telluride (PbTe) is one of the most efficient thermoelectric materials from 400~K to 800~K, owing to its low lattice thermal conductivity, as well as high electrical conductivity and Seebeck coefficient when appropriately doped~\cite{Heremans2008,Pei2011,LaLonde2011a,Pei2011a}. Nevertheless, it is still unclear which scattering mechanisms determine electronic transport in PbTe. In some previous works~\cite{Allgaier1958,Lalonde2011,Pei2014}, 
the main scattering mechanism has been attributed to acoustic phonons. Other studies argued that scattering due to polar and non-polar optical phonons is also important at certain temperatures and carrier concentrations~\cite{Ravich1971a,Zayachuk1997,Freik2002,Bilc2006,Vineis2008}. In all these studies, the parameters related to the strength of scattering due to acoustic and non-polar optical phonons (acoustic and optical deformation potentials, respectively) were determined empirically, by fitting their values to electronic transport measurements. However, our recent first principles calculations have shown that the widely used empirical values of the acoustic deformation potentials of $n$-type PbTe are largely overestimated~\cite{Murphy2018}. Consequently, the relative strength of different electron-phonon (e-ph) scattering mechanisms and their contribution to electronic transport in $n$-type PbTe is still unknown.   

It is of particular interest to understand the strength of e-ph scattering due to transverse optical (TO) phonons since they play a major role in establishing the low lattice thermal conductivity of PbTe. Inelastic neutron scattering measurements~\cite{Delaire2011} and first-principles calculations~\cite{An2008} have shown that the TO modes near the zone center are soft and have very small frequencies ($\sim 1$~THz). They are strongly coupled to heat carrying acoustic phonons, and lead to low lattice thermal conductivity~\cite{Shiga2012,Romero2015,Murphy2016,Murphy2017}. These properties raise the question about the strength of the interaction between these soft TO modes and the electronic states near the conduction band minima at L. If this interaction were weak, it would mean that soft TO modes are the key to the high thermoelectric figure of merit of PbTe: they preserve its high electronic conductivity while suppressing the lattice thermal conductivity. 

In recent years, it has become possible to calculate the strength of e-ph scattering of semiconductors  in the entire Brillouin zone (BZ) purely from first principles~\cite{Giustino2007,Sjakste2015,Verdi2015,Gunst2016,Giustino2017}. Methods for solving the linearized Boltzmann transport equation (BTE) and calculating the electronic transport coefficients from first principles have also been developed~\cite{Li2015,Bonini2016,Chen2017,Ponce2018}. However, these methods suffer from a large computational cost. This is due to the fact that electronic conduction occurs only in a small energy window close to the Fermi level. In the momentum space, this energy window corresponds to a very small fraction of the BZ. To converge transport coefficient values, a high sampling density is necessary to ensure that enough sampling points fall into this critical volume. In many first principles calculations~\cite{Li2015,Bonini2016,Chen2017}, the whole BZ is sampled brute-force, using a very dense mesh. On the other hand, since the energy window of interest is small, using a simpler BTE model with parameters calculated from first principles (e.g. effective masses, deformation potentials) may offer an efficient and accurate alternative to describe electronic transport in semiconductors.
 
It is especially challenging to accurately calculate from first principles the electronic transport properties of direct narrow-gap semiconductors, such as PbTe. The reason for this is the well-known band gap underestimation using the standard approximations for the exchange-correlation energy in density functional theory (DFT), such as the local density approximation (LDA)~\cite{Perdew1985}. It has been shown that the LDA with spin-orbit coupling (SOC) leads to the incorrect character of the states near the band gap in PbTe i.e.~the valence and the conduction bands are inverted and the resulting band gap is negative~\cite{Hummer2007,Svane2010,Murphy2018}. As a result, the effective masses of PbTe computed using this approach are in large disagreement with experiments. Furthermore, the calculated parameters for e-ph scattering (deformation potentials, dielectric constants) may also not be reliable~\cite{Murphy2018}. Recently, such electronic band structure has been used to calculate the thermoelectric transport coefficients of PbTe from first principles~\cite{Song2017}. More accurate electronic band structures of PbTe were used only in conjunction with the constant relaxation time approximation in the Seebeck coefficient calculations~\cite{Singh2010,Parker2013,Berland2018}. In contrast to the LDA including SOC, both the hybrid HSE03 exchange-correlation functional including SOC and the LDA excluding SOC correctly describe the curvature of the states near the band gap~\cite{Hummer2007,Svane2010,Murphy2018}. The parameters obtained from these accurate band structure calculations can be readily used in charge transport models, and allow us to identify  the main electronic scattering mechanisms in PbTe.

In this paper, we present an electronic transport model developed for $n$-type PbTe based on the Boltzmann transport equation within the transport relaxation time approximation. All the input parameters are calculated \emph{ab-initio}, using the exchange-correlation functionals that correctly describe the character of the electronic states close to the band edges (the hybrid HSE03 exchange-correlation functional including SOC and the LDA excluding SOC). We find that electron-longitudinal optical (LO) phonon scattering dominates electronic transport in $n$-type PbTe over a range of temperatures (up to $300$~K) and carrier concentrations (up to $\sim$10$^{20}$~cm$^{-3}$), due to weak screening effects. We show that scattering due to acoustic phonons is much weaker than previously assumed~\cite{Ravich1971a,Zayachuk1997,Pei2014}. Only at very high concentrations, acoustic and LO phonon scattering become comparable. Using a symmetry analysis, we show that the e-ph matrix elements for the electronic states at the conduction band minimum and the soft TO phonons at $\Gamma$ are zero. Our detailed first principles calculations confirm that electron-TO phonon scattering is indeed the weakest scattering mechanism in PbTe. This finding elucidates the central role of soft TO modes and their weak interaction with conducting states in establishing the high thermoelectric figure of merit of PbTe. 

%%%%%%%%%%%%%%%%%%%%%%%%%%%%%%%%%%%%%%%%%%
%%   Methods for transport calculations
%%%%%%%%%%%%%%%%%%%%%%%%%%%%%%%%%%%%%%%%%%
\section{Method}
\label{sec:method}

%------------------------------------------------------------
%  Boltzmann transport equation
%------------------------------------------------------------
\subsection{Boltzmann transport equation in the transport relaxation time approximation}
\label{ssec:boltzmann-transport}

Under a weak electrical field, solving the Boltzmann transport equation in the steady-state gives the electrical conductivity tensor~\cite{Ziman1960a}
\begin{equation}
\label{eq:cond}
    \sigma^{ij} = - \frac{2 e^2}{V N_{\kv}} \sum_{\one}  \pdd{f^0_\one}{E_\one} \tau_{\one} v^i_{\one} v^j_{\one},
\end{equation}
where $e$ is the electron charge, $V$ is the unit cell volume, $N_{\kv}$ is the number of \kv\ points sampled in the first BZ, $i$ and $j$ are the Cartesian directions, and $\vv_\one$ and $f^0_\one$ are the group velocity and the equilibrium Fermi-Dirac occupation of an electronic state with the crystal momentum \kv\ in the band $n$. The deviation of the distribution function from $f^0_\one$ due to the electrical field is encoded in the transport relaxation time $\tau_{\one}$. Electronic drift mobility can be obtained as $\bm{\mu} = \bm{\sigma} / e n_c$, where $n_c$ is the carrier concentration at temperature $T$, given by $n_c = \frac{2}{VN_{\kv}
} \sum_{\one} f^0_\one$.

The transport relaxation time $\tau_\one$ is given by~\cite{Gunst2016,Sohier2014}
\begin{equation}
 \label{eq:TRTA}
 \tau^{-1}_\one = \sum_{m \qv} \Pot \frac{1-f^0_\two}{1-f^0_\one}
  \left(1 -  \beta_\one^\two \right),
\end{equation}
with $\Pot$ denoting the transition rate from initial $\one$ to final state $\two$, and $\beta_\one^\two = \bm{v}_\one \cdot \bm{v}_\two/(\abs{\bm{v}_\one}\abs{\bm{v}_\two})$ characterizing the scattering angle. 
We note that the last term in Eq.~\eqref{eq:TRTA} is sometimes neglected in the literature~\cite{Zhou2016,Song2017,Restrepo2014}. This approximation corresponds to using the inverse of the imaginary part of the e-ph self-energy i.e. $\tau_\one = (2/\hbar \Im \Sigma_\one)^{-1}$. We refer to this approximation as the self-energy relaxation time (SERT) approximation, while Eq.~\eqref{eq:TRTA} represents the transport relaxation time  (TRT) approximation. We will show explicitly that the SERT approximation gives much underestimated results for the electronic mobility of $n$-type PbTe compared to the TRT approximation. On the other hand, the TRT is usually a very good approximation for the full numerical solution of the linearized BTE~\cite{Ma2018,Sohier2014}.
 
In this work, we consider electron-phonon and ionized-impurity scattering. For e-ph scattering, the scattering term \Pot\ in Eq.~\eqref{eq:TRTA} can be written as~\cite{Ridley1999c}
\begin{equation}
\begin{split}
\label{eq:rate}
\Pot = \frac{2\pi}{\hbar} \sum_{\lambda, \pm} & \abs{g_{\one,\qv \lambda}^{\two}}^2 \left(N^0_{\qv \lambda}+\frac{1}{2}\mp \frac{1}{2} \right) \\
   & \times \delta (E_{\one} \pm \hbar \omega_{\qv \lambda} - E_{\two}),
\end{split}
\end{equation}
where $\hbar$ is the reduced Planck constant, and $E_\one$ is the electron energy. $N^0_{\qv \lambda}$ and $\omega_{\qv \lambda}$ are the equilibrium distribution and the frequency of a phonon with the crystal momentum \qv\ and the branch $\lambda$, and $g_{\one,\qv \lambda}^{\two}$ is the e-ph Hamiltonian matrix-element. The upper/lower sign in $\pm$ and $\mp$ in the energy conservation denotes phonon absorption/emission. 

We parametrize $E_\one$, $\omega_{\qv \lambda}$ and the matrix-elements $g_{\one,\qv \lambda}^{\two}$ for each phonon mode using first-principles calculations, as described in the following subsections. This enables us to exactly solve the energy conservation condition in Eq.~\eqref{eq:rate}, see Appendix~\ref{apx:kane}. In contrast, many first principles calculations~\cite{Song2017,Gunst2016,Fiorentini2016} evaluate the delta function in Eq.~\eqref{eq:rate} using a Lorentzian or Gaussian with an empirical parameter for energy broadening, or using the linear tetrahedron method~\cite{Lehmann1972}. Our parametrization and the exact energy conservation enable us to use a very fine \kv\ point sampling in those parts of the BZ which contribute most to mobility, and thus substantially reduce the computational cost compared to the standard first-principles methods~\cite{Song2017,Gunst2016,Fiorentini2016}. 

%------------------------------------------------------------
%  Electronic bands 
%------------------------------------------------------------
\subsection{Electronic band structure}
\label{ssec:elec-band}

PbTe is a direct narrow-gap semiconductor whose gap is located at four equivalent L points. The conduction band L valleys (CB$_{\rm L}$) are far below in energy from other conduction band minima \cite{Murphy2018}. Consequently, electrons will partially occupy only the L valleys for a large range of temperatures and doping concentrations, and only the L valleys will contribute to electronic conduction. Such electronic band structure is usually well described using the Kane model derived from the \kdotp\ theory~\cite{Kane1957,RAVIC1971}.

The main assumption of the Kane model is that the top valence and bottom conduction bands are strongly coupled to each other, and weakly coupled to all other bands. Due to the strong coupling between the valence and conduction bands, the energy dispersion curve in the Kane model is non-parabolic and for the band minimum at the L point in the [111] direction reads as
\begin{equation}
\label{eq:gkane-model}
\frac{\hbar^2}{2} \left( \frac{\kpara^2}{\mpara} + \frac{\kperp^2}{\mperp} \right)
=  E \left( 1 + \alpha E \right), 
\end{equation}
where $\alpha$ is the non-parabolicity parameter, \kpara\ and \kperp\ are the components of the wavevector parallel and perpendicular to the [111] direction, and \mpara\ and \mperp\ are the parallel and perpendicular effective masses. If the coupling of the top valence and bottom conduction bands with all other bands is neglected, then $\alpha=1/E_g$, where $E_g$ corresponds to the direct band gap~\cite{Kane1957}. We will refer to this approximation as the Kane 2-band model, while the more general case with $\alpha\ne 1/E_g$ will be denoted as the generalized Kane model. Analytical expressions for all the quantities that enter Eqs.~\eqref{eq:cond}, \eqref{eq:TRTA} and \eqref{eq:rate} obtained from the Kane model are given in Appendix~\ref{apx:kane}. We will verify the applicability of both the generalized and 2-band Kane models by comparison with the first-principles electronic band structures that correctly capture the character of the electronic states near the band gap: the hybrid HSE03 functional including SOC and the LDA excluding SOC.

We use the Vienna \emph{ab-initio} simulation package (VASP)~\cite{KRESSE1996} to perform electronic band structure calculations using the screened Heyd-Scuseria-Ernzerhof (HSE03) hybrid functional \cite{Heyd2003,Heyd2004} including the spin-orbit coupling (SOC). The basis set for the one-electron wave functions is constructed with the Projector Augmented Wave (PAW) method~\cite{Kresse1999}. For the PAW pseudopotentials, we include the $5d^{10}6s^2 6p^2$ states of Pb and $5s^2 5p^4$ states of Te as the valence states. A cutoff energy of 18.4 eV and a 8$\times$8$\times$8 \kv-mesh are used to calculate the electronic band structure of PbTe. 

We also carry out DFT calculations with the {\sc ABINIT} code~\cite{Gonze2009,Gonze2016} using the LDA~\cite{Caperley1980,Perdew1981} excluding SOC. Here, we turn off the SOC to ensure a positive band gap and physically correct conduction and valence band states near the L point in PbTe. We use the Hartwigsen-Goedecker-Hutter (HGH) norm-conserving pseudopotentials with the $6s^2 6p^2$ states of Pb and $5s^2 5p^4$ states of Te explicitly included as the valence states~\cite{Hartwigsen1998}. The electronic band structure is calculated using a cutoff energy of 45 Ha and a four-shifted 12$\times$12$\times$12 \kv-mesh. The LDA without SOC is also used to calculate all deformation potential values, phonon frequencies, dielectric and elastic constants using density functional perturbation theory (DFPT)~\cite{Gonze1997,Baroni2001} and the {\sc ABINIT} code.

%------------------------------------------------------------
%  Scattering mechanisms
%------------------------------------------------------------
\subsection{Scattering mechanisms}
\label{ssec:scattering-terms}

Since electronic conduction in $n$-type PbTe occurs only through the L valleys, we consider scattering  within an L valley via phonons near the $\Gamma$ point, as well as scattering between non-equivalent L valleys via phonons near the X point. Our treatment of these scattering mechanims from first principles will be described in this subsection, where we give explicit expressions for the e-ph matrix-elements for each phonon mode. We also consider ionized-impurity scattering, whose details are given in Appendix~\ref{apx:ionized-impurity}. The transport relaxation times (TRTs) of different scattering channels are combined via Matthesien's rule to find the total TRTs and the electronic mobility of PbTe via Eq.~\eqref{eq:cond}. 

%------------------------------------------------------------
%  Symmetry-forbidden scattering mechanisms
%------------------------------------------------------------
\subsubsection{Symmetry-forbidden scattering mechanisms}
\label{sssec:crystal-symm}

We first show that the e-ph matrix elements of PbTe corresponding to scattering of the CB state at the L point via a $\Gamma$ point phonon, as well as scattering between the CB states at two inequivalent L points via an X point phonon, are zero. In PbTe, the center of inversion is the Pb or Te site, and thus all X and $\Gamma$ point phonons have odd parity under inversion symmetry~\cite{Dresselhaus2008}. Odd parity phonons can only couple between electronic states of opposite parity since e-ph matrix elements are invariant under symmetry operations~\cite{Dresselhaus2008}. The e-ph matrix-elements $\mel{\psi_k}{H_{ep}}{\psi_{k+\Gamma}}$ and $\mel{\psi_k}{H_{ep}}{\psi_{k+X}}$ thus vanish exactly at the L point since initial and final electronic states are of the same parity. Our direct calculations of these matrix elements obtained using DFPT-LDA also confirm that these matrix elements are zero.

Based on the symmetry analysis above, we neglect intervalley scattering via long wavevector phonons near the X point in the rest of the paper. The comparison of our calculated scattering rates with detailed DFPT-LDA e-ph calculations outside of the high symmetry points shows that this is a reasonable approximation (see Section~\ref{ssec:el-ph-scattering}). For short wavevector phonons near the $\Gamma$ point, we take into account the linear dependence on \qv\ of the corresponding e-ph matrix-elements, as described in the following.

%------------------------------------------------------------
%  Acoustic phonon scattering
%------------------------------------------------------------
\subsubsection{Intravalley acoustic phonon scattering}
\label{apx:acoustic-phonon}

To describe scattering due to acoustic phonons in the long-wavelength limit, we use a generalized deformation potential theory developed by Herring and Vogt for anisotropic many-valley semiconductors~\cite{Herring1956}. For a cubic material with the conduction band minima at the L valleys, the interaction between the CB$_{\rm L}$ electrons and acoustic phonons for $\qv \rightarrow 0$ can be described as~\cite{Murphy2018}
\begin{equation}
\label{eq:hep}
H_{ep}  =  \Xi^{\text{ac}}_d {\bf e}_{\qv \lambda} \cdot {\bf q} + \Xi^{\text{ac}}_u \left( {\bf e}_{\qv \lambda} \cdot \hat{\bf k}_{\text{L}} \right) \left( {\bf q} \cdot \hat{\bf k}_{\text{L}} \right),
\end{equation}
where \Dad\ and \Dau\ are the two linearly independent elements of the acoustic deformation potential tensor, ${\bf e}_{\qv \lambda}$ is the strain polarization vector, and $\hat{{\bf k}}_{\text{L}}$ is the unit vector parallel to the ${\bf k}$-vector of the L valley. The dilatation deformation potential $\Dad$ represents the band shift due to a dilatation in the two directions normal to the symmetry axis of the L valley~\cite{Herring1956}. The uniaxial deformation potential $\Dau$ corresponds to the band shift due to a uniaxial shear along the symmetry axis of the L valley~\cite{Herring1956}.

Following the procedure outlined by Herring and Vogt~\cite{Herring1956}, we obtain the expressions for the electron-acoustic phonon matrix elements starting from the deformation potential Hamiltonian given by Eq.~\eqref{eq:hep}. Since acoustic phonon frequencies are small, we use $N^0_{\qv \lambda}\approx k_BT/\hbar\omega_{\qv \lambda}$ in Eq.~\eqref{eq:rate} for both absorption and emission processes. If we define $M^\lambda \equiv |g_{\kv,\qv \lambda}^{\kv+\qv}|^2 \times N^0_{\qv \lambda}$ and assume that phonon frequencies change linearly with \qv\ , these terms for longitudinal and transverse acoustic phonons can be written as~\cite{Herring1956}
\begin{equation}
\label{eq:gkk-la}
\begin{cases}
M^{\rm LA} = \frac{k_B T  I_{\kv,\kv+\qv}^2}{V} (\Dad + \Dau \cos^2 \theta)^2  \alpha_l,   \\
M^{\rm TA1}+M^{\rm TA2} = \frac{k_B T I_{\kv,\kv+\qv}^2 }{V}  (\Dau)^2 \cos^2\theta \sin^2\theta \alpha_t, 
\end{cases}
\end{equation}
where $V$ is the unit cell volume, $\theta$ is the angle between $\qv$ and the parallel direction of the $L$ valley, and $I_{\kv,\kv+\qv}=\braket{\kv}{\kv+\qv}$ is the overlap integral between the two Bloch wave-functions. Also, $\alpha_l=[1+(2/3 c^* (0.15-1.50 \cos^2 \theta +1.75 \cos^4 \theta)/(c_{12}+2c_{44}+\frac{3}{5}c^*)] /(c_{12}+2c_{44}+\frac{3}{5}c^*)$ and $\alpha_t = [0.375/c_{44} + 0.625/(c_{44}+c^*/3) + 9/8\times \cos^2 \theta(1/c_{44} - 1/(c_{44}+c^*/3)]$, where $c_{11}$, $c_{12}$ and $c_{44}$ are the elastic constants and $c^*=c_{11}-c_{12}-2c_{44}$~\cite{Herring1956}. We neglect acoustic phonon energies when calculating the scattering rate given by Eq.~\eqref{eq:rate}.

We calculate deformation potentials \Dad\ and \Dau\ by fitting the e-ph matrix elements obtained using DFPT with the deformation potential Hamiltonian of Eq.~\eqref{eq:hep} in the limit of ${\bf q}\rightarrow 0$, as explained in detail in Ref.~\onlinecite{Murphy2018}. The calculated values of \Dad\ and \Dau\ are given in Table~\ref{tab:parameters}, together with our DFPT values of elastic constants and our DFT value of the lattice constant $a_0$.

%------------------------------------------------------------
%  Optical phonon scattering
%------------------------------------------------------------
\subsubsection{Intravalley optical phonon scattering}
\label{apx:optical-phonon}

As already discussed, the zero-order terms of the matrix elements between the CB$_{\rm L}$ states and the zone center optical phonons are zero in PbTe. To account for the short-range (non-polar) part of this interaction, we consider the first-order terms of these matrix elements~\cite{Harrison1956,Ferry1976}. Due to symmetry reasons, the e-ph interaction Hamiltonian for non-polar optical phonons has the same form as that for acoustic phonons~\cite{Harrison1956,Ferry1976}:
\begin{equation}
\label{eq:hep2}
H_{ep}  =  \Xi^{\text{opt}}_d {\bf \bar{e}}_{\qv \lambda} \cdot {\bf q} + \Xi^{\text{opt}}_u \left( {\bf \bar{e}}_{\qv \lambda} \cdot \hat{\bf k}_{\text{L}} \right) \left( {\bf q} \cdot \hat{\bf k}_{\text{L}} \right),
\end{equation}
where \Dod\ and \Dou\ are the two linearly independent elements of the optical deformation potential tensor, and ${\bf \bar{e}}_{\qv \lambda}$ is the unit vector in the direction of atomic displacements.

The expressions for the matrix elements due to the non-polar interaction with longitudinal and transverse optical phonons can be derived similarly to those for acoustic phonons~\cite{Harrison1956,Ferry1976}, and read as
\begin{equation}
\label{eq:gkk-lo}
\begin{cases}
|g^{\rm LO}_{\rm DP}|^2 = \frac{\hbar q^2 I_{\kv,\kv+\qv}^2}{2M \omega^{\rm LO}_\qv} (\Dod +\Dou \cos^2\theta )^2, \\
|g^{\rm TO1}_{\rm DP}|^2+|g^{\rm TO2}_{\rm DP}|^2 =  \frac{\hbar q^2  I_{\kv,\kv+\qv}^2}{2M\omega^{\rm TO}_\qv}  (\Dou)^2 \cos^2 \theta 
 \sin^2 \theta, 
\end{cases}
\end{equation}
where $M$ is the mass of the unit cell, and $\omega^{\rm LO}_\qv$ and $\omega^{\rm TO}_\qv$ are the LO and TO frequencies. Owing to the finite frequency of optical phonons, we distinguish explicitly between the absorption and emission processes in the scattering rate given by Eq.~\eqref{eq:rate}, and use the Bose-Einstein distribution.  

The optical deformation potentials \Dod\ and \Dou\ are obtained from first principles, in the same manner as the acoustic deformation potentials using DFPT-LDA calculations~\cite{Murphy2018}. Their values are given in Table~\ref{tab:parameters}. The \qv\ dependence of the TO phonon frequency is modeled by a quadratic function for $\qv \rightarrow 0$ ($\omega^{\rm TO}_\qv=\omega^{\rm TO}_{\Gamma}+\partial^2\omega^{\rm TO}/\partial |\qv|^2 \times |\qv|^2$) and fitted to the DFPT-LDA calculations along $\Gamma-X$ and $\Gamma-K$ directions.
For LO mode, we consider a constant phonon frequency equal to our DFPT value at the $\Gamma$ point, which is a reasonable approximation of its dispersion~\cite{Murphy2016}. The DFPT values of $\omega^{\rm TO}_{\Gamma}$, $\partial^2\omega^{\rm TO}/\partial |\qv|^2$ and $\omega^{\rm LO}_{\Gamma}$ are listed in Table~\ref{tab:parameters}.

In addition to the non-polar interaction, the atomic displacements corresponding to the LO mode of PbTe create long-range interactions between electrons and LO modes. The long-range interaction can be described by the Fröhlich model~\cite{Frohlich1954}, whose e-ph matrix element is given by
\begin{equation}
\label{eq:gkk-froh}
|g^{\rm LO}_{\rm F}|^2 = \frac{\hbar e^2 \omega^{\rm LO}_\qv}{2 V \epsilon_0} \left( \frac{1}{\epsilon_\infty} - \frac{1}{\epsilon_s} \right)\frac{q^2  I^2_{\kv,\kv+\qv} }{\left[q^2 + (q_{\rm scr}^{\infty})^2\right]^2} ,
\end{equation}
where $\epsilon_0$ is the vacuum permittivity, and $\epsilon_\infty$ and $\epsilon_s$ are the high-frequency and static dielectric constant whose DFPT values are given in Table~\ref{tab:parameters}. The effect of screening by mobile charges is included via the screening wavevector $q_{\rm scr}^{\infty}$, which is taken into account using the Thomas-Fermi model~\cite{Ridley1999b} (see Appendix~\ref{apx:screening}). The short- and long-range contributions to the electron-LO phonon interaction are coherently combined as $g^{\rm LO} =  g^{\rm LO}_{\rm F} + g^{\rm LO}_{\rm DP}$.

%%%%%%%%%%%%%%%%%%%%%%%%%%%%%%%%%%%%%%%%%%
%%   Results
%%%%%%%%%%%%%%%%%%%%%%%%%%%%%%%%%%%%%%%%%%
\section{Results and discussion}
\label{sec:results}

\begin{table}[t!]
    \begin{tabularx}{0.9\columnwidth}{X >{\hsize=.5\hsize}X >{\hsize=.6\hsize}X}
        \hline
        \hline
        Parameter   &  Value    &   Experiment            \\
        \hline
        \Dad\ (eV)  &   0.37    &  12-22$^{a}$  \\
        \Dau\ (eV)  &   7.03    &  4.5$^b$        \\
        \Dod\ (eV)  &  19.09    &  - \\
        \Dou\ (eV)  & -34.24    &  - \\
        \hline
$\omega^{\rm LO}_{\Gamma}$ (THz) &   3.2    &    3.42$^{c,f}$\\
$\omega^{\rm TO}_{\Gamma}$ (THz) &   1.0    &    0.95$^{c,f}$\\
   $\partial^2\omega^{\rm TO}/\partial |\qv|^2$ (THz \AA$^{2}$) & 15  & -\\
        \hline
        $\epsilon_s$        &   356.8 & 478$^c$, 412$^f$ \\
        $\epsilon_\infty$   &  34.85& 36.9$^c$, 31.81$^f$ \\
        \hline
        $c_{11}$ (GPa)  &   136.4    & 128.1$^d$, 107.2$^f$  \\
        $c_{12}$ (GPa)  &   3.8      & 4.4$^d$, 7.68$^f$  \\
        $c_{44}$ (GPa)  &   17.1     & 15.14$^d$, 13.0$^f$ \\
        \hline
        $a_0$ (nm)   &    0.634      & 0.642$^e$  \\        
        \hline
        \mpara/$m_0$        &   0.216   &   0.24$^c$, 0.21$^g$\\
        \mperp/$m_0$        &   0.037   &   0.024$^c$,0.021$^g$\\
        \Eg (eV)            &   0.237   &   0.19$^c$ \\
        $\partial \Eg / \partial T$ ($10^{-4}$eV/K) &  2.90 & 3.2$^e$ \\
        $\alpha$ (eV$^{-1}$) &   2.0   &     -      \\
        \hline
        \hline
    \end{tabularx}
    \begin{flushleft}
    $^a$Obtained by fitting a model to electronic transport measurements from Ref.~\onlinecite{Pei2014,RAVIC1971,Zayachuk1997}.\\
    $^b$Ref.~\onlinecite{RAVIC1971}, 
    $^c$Ref.~\onlinecite{DALVEN1974}, 
    $^d$Ref.~\onlinecite{Zasavitskii2004},
    $^e$Ref.~\onlinecite{Gibbs2013},
    $^f$Ref.~\onlinecite{Cochran1966},
    $^g$Ref.~\onlinecite{Pascher2003}. \\
    \end{flushleft}
    \caption{Parameters used in the calculation of the electronic mobility of $n$-type PbTe, computed from first principles: acoustic deformation potentials (\Dau\ and \Dad, see Eq.~\eqref{eq:gkk-la}), optical deformation potentials (\Dou\ and \Dod, see Eq.~\eqref{eq:gkk-lo}), optical phonon frequencies ($\omega^{\rm LO}_{\Gamma}$, $\omega^{\rm TO}_{\Gamma}$, and $\partial^2\omega^{\rm TO}/\partial |\qv|^2$, see Eq.~\eqref{eq:gkk-lo}), static and high-frequency dielectric constant ($\epsilon_s$ and  $\epsilon_\infty$, see Eqs.~\eqref{eq:gkk-froh}, \eqref{eq:gkk-ii} and \eqref{eq:qscrn}), elastic constants ($c_{11}$, $c_{12}$ and $c_{44}$, see Eq.~\eqref{eq:gkk-la}), lattice constant ($a_0$), parallel and perpendicular effective masses (\mpara\ and \mperp, see Eq.~\eqref{eq:gkane-model}), direct band gap and its temperature coefficient, and non-parabolicity parameter ($\alpha$, see Eq.~\eqref{eq:gkane-model}).    }
    \label{tab:parameters}
\end{table}

\subsection{Electronic band structure}
\label{ssec:el-bands}

\begin{figure}[t!]
    \centering
    \includegraphics[width=\columnwidth]{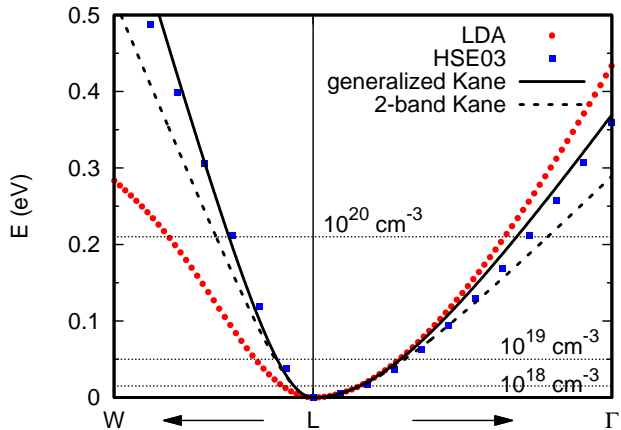}
    \caption{The conduction band of PbTe near the L point at 0~K calculated from first principles using the local density approximation (LDA) without spin-orbit coupling (red dots) and the HSE03 hybrid functional with spin-orbit coupling (blue squares). The conduction band minima from the LDA and HSE03 calculations are aligned. The path L-$\Gamma$ and L-W corresponds to the parallel and perpendicular direction, respectively. The fits to the HSE03 band using the generalized and 2-band Kane model are shown in solid and dashed black lines, respectively. Fermi levels for different doping concentrations calculated with the HSE03 bands are also indicated by dotted black lines. }
    \label{fig:CB-compare-with-model}
\end{figure}

We first discuss the electronic band structure of $n$-type PbTe obtained from first principles. Fig.~\ref{fig:CB-compare-with-model} shows the conduction band near the minimum at the L point in the directions parallel and perpendicular to the [111] direction, calculated using the HSE03 functional including SOC and the LDA excluding SOC. We also illustrate the Fermi levels for the doping concentrations of $\bar{n}=10^{18}$~cm$^{-3}$, $10^{19}$~cm$^{-3}$ and $10^{20}$~cm$^{-3}$ calculated with the HSE03 bands. The band structure obtained with the LDA excluding SOC compares well with that of the hybrid functional for the doping concentrations of $\sim 10^{19}$~cm$^{-3}$ and lower in the parallel direction. On the other hand, the LDA conduction band deviates from that of the HSE03 calculation along the perpendicular direction. Previous works~\cite{Hummer2007,Murphy2018} show that the HSE03 functional yields the values of band gap and effective masses that are in very good agreement with experiment~\cite{Dalven1969,Pascher1984} and {\sc GW} calculations~\cite{Svane2010}. Our LDA excluding SO calculation gives a similar value for \mpara\ as the HSE03 calculation and experiment, and overestimates \mperp\ ($\sim 25$\% larger than HSE03 and experiment)~\cite{Murphy2018}. We also showed that the HSE03 with SOC and the LDA without SOC predict similar values of the acoustic deformation potentials for $n$-type PbTe~\cite{Murphy2018}. Consequently, here we use the HSE03 electronic structure to obtain the parameters for the Kane model, while all other parameters are calculated using the LDA without SOC.  

 We find that we can accurately fit the conduction band near the L point calculated with the hybrid functional using the Kane model if we use the values of the effective masses calculated using the LDA excluding SOC as the parameters of the Kane model (\mpara = 0.216$m_e$ and \mperp = 0.037$m_e$). These fits are shown in Fig.~\ref{fig:CB-compare-with-model}. The generalized Kane model with $\alpha$ = 2.0~eV$^{-1}$ accurately reproduces the HSE03 conduction band in a large energy window, even for doping concentrations of $10^{20}$~cm$^{-3}$. The band obtained using the Kane 2-band model with $\alpha=1/E_g=4.22$~eV$^{-1}$ shows a slightly worse agreement with the HSE03 calculation in the perpendicular direction. Even though $\alpha=2.0$ eV$^{-1}$ gives a more accurate description of the conduction band L valley at $0$~K compared to $\alpha=4.22$ eV$^{-1}$, in this paper we use both values to evaluate the sensitivity of our mobility values to electronic band variations. Unless explicitly stated, we use $\alpha=2.0$ eV$^{-1}$ in our calculations. We note that the fitting is done in a wide energy window instead close to the band minimum to represent well a wide range of doping concentrations (up to $10^{20}$~cm$^{-3}$). Therefore, our set of parameters is not expected to work ideally at very low carrier density and low temperature.

We also account for the temperature dependence of the electronic band structure of PbTe by calculating the temperature variation of the band gap from first principles. We computed the temperature dependence of $E_g$ accounting both for the contributions of thermal expansion and e-ph coupling~\cite{Querales2018}, where the latter effect was computed using the Allen-Heine-Cardona approach~\cite{Allen1976,Allen1981,Allen1983} and DFPT~\cite{Ponce2015}. Our calculated $\partial E_g / \partial T=2.9\times 10^{-4}$~eV/K is in very good agreement with experiment~\cite{Gibbs2013}. The temperature dependence of the gap leads to the temperature dependent non-parabolicity parameter and effective masses in the Kane 2-band model ($m^*(T)/m^*(0~\text{K})=\alpha(0~\text{K})/\alpha(T)=E_g(T)/E_g(0~\text{K})$)~\cite{Dresselhaus2008}. In the generalized Kane model, we assume that the temperature dependence of $\alpha$ and $m^*$ is the same as in the Kane 2-band model even though $\alpha\ne 1/E_g$. This approximation may be reasonable in the temperature range considered here ($100\--300$~K), where the temperature variations of $\alpha$ and effective masses have a small effect on mobility. 

\subsection{Electron-phonon scattering rates}
\label{ssec:el-ph-scattering}

\begin{figure}[t!]
    \centering
    \includegraphics[width=\columnwidth]{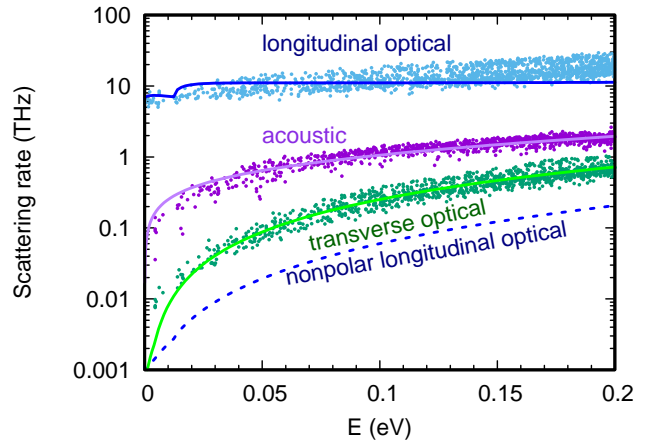}
    \caption{Electron-phonon scattering rates of $n$-type PbTe for different phonon modes versus electronic energy. Dots represent the scattering rates calculated using the electron-phonon Wannier approach and density functional perturbation theory (see text for explanation), and solid lines correspond to the scattering rates computed using our model parameterized from first principles. Dashed blue line shows the non-polar contribution to longitudinal optical scattering from our model. }
    \label{fig:scat-rates}
\end{figure}

We next verify our model by comparing the e-ph scattering rates with those computed from first principles using a general form of the e-ph Hamiltonian (see Appendix~\ref{apx:el-ph}) and Eq.~\eqref{eq:rate}, without using any assumptions of the deformation potential theory. The e-ph matrix elements were first calculated using DFPT as implemented in {\sc Quantum ESPRESSO}~\cite{QE-2017, QE-2009} with the LDA excluding SOC using 10$\times$10$\times$10 \kv\ and \qv\ grids and a cutoff energy of 45~Ha. We then used the {\sc EPW} code\cite{Ponce2016} to interpolate these matrix elements on finer 80$\times$80$\times$80 \kv\ and \qv\ grids using a real-space Wannier functions approach~\cite{Giustino2007,Giustino2017}. In these calculations, we use the norm-conserving fully relativistic pseudopotentials with the $6s^2 6p^2$ states of Pb and $5s^2 5p^4$ states of Te as the valence states \cite{pseudo}. The electronic band structure and the e-ph matrix elements of PbTe calculated in this way using {\sc Quantum ESPRESSO} are very similar to those calculated with the {\sc ABINIT} code using the LDA excluding SOC and HGH pseudopotentials. In the EPW calculation, we used the broadening parameter of 30~meV for the energy conservation. The screening effect is not considered in these calculations. The scattering rates for $n$-type PbTe at 300~K due to different phonon modes are illustrated in Fig.~\ref{fig:scat-rates}. The very good agreement between the scattering rates computed with our model (solid lines) and the EPW approach (dots) confirms the validity and accuracy of our model.

Our results show that LO phonon scattering is the dominant scattering mechanism in $n$-type PbTe with low carrier concentrations. Scattering due to polar long-range interactions represents the main contribution to LO scattering, while the short-range contribution is very small, as shown by the dashed blue line in Fig.~\ref{fig:scat-rates}. The polar nature of LO phonon scattering can also be observed from the weak energy dependence of the scattering rate. If the electron energy is smaller than the LO frequency, LO phonons can only be absorbed. When the electron energy becomes higher than the LO frequency, emission processes steeply increase the LO phonon scattering rate.

We also find that acoustic phonon scattering is one order of magnitude weaker than LO scattering for low energy electrons, as shown in Fig.~\ref{fig:scat-rates}. Acoustic scattering rate has a larger energy dependence than that due to LO phonons (proportional to the electronic density of states), which makes it comparable to LO scattering for higher energies. The relatively weak acoustic scattering can be understood from the fact that our first principles value of the acoustic dilatation deformation potential (\Dad=0.37~eV) is substantially lower than those obtained by fitting electronic transport measurements ($\Dad\sim$12-22~eV)~\cite{Ravich1971a,Zayachuk1997,Pei2014}, as discussed in Ref.~\onlinecite{Murphy2018}.

Scattering due to soft TO phonons is by far the weakest e-ph scattering mechanism, owing to the fact that TO scattering vanishes by symmetry for the conduction band state at L and the zone center TO modes. This finding suggests that soft TO modes do not degrade electronic transport of $n$-type PbTe, while strongly suppressing its lattice thermal conductivity. We note that our calculated optical deformation potentials given in Table~\ref{tab:parameters} are of similar order of magnitude to those obtained empirically ($\Xi^{\text{opt}}\sim 15\--32$~eV)~\cite{Zayachuk1997,Bilc2006,Vineis2008}. Finally, our conclusions about the relative strength of each scattering mechanism qualitatively agree with those of Ref.~\onlinecite{Song2017}, in spite of different electronic band structures.

\begin{figure}[t!]
    \centering
    \includegraphics[width=\columnwidth]{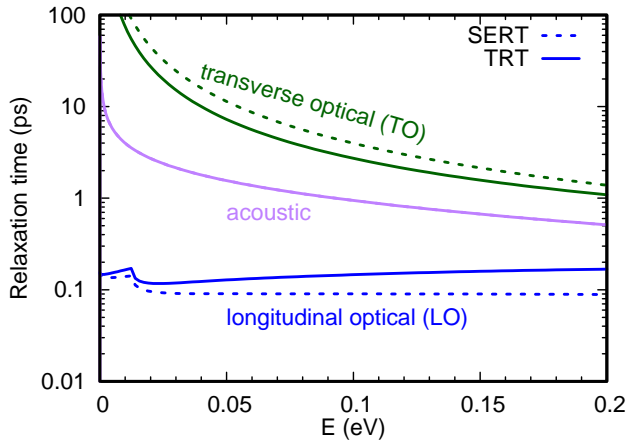}
    \caption{Relaxation times of $n$-type PbTe for different phonon modes versus electronic energy. Solid lines represent transport relaxation times (see text for explanation), while dashed lines correspond to self-energy relaxation times i.e. inverse scattering rates. The two relaxation times are identical for acoustic phonons.}
    \label{fig:sr-vs-rt}
\end{figure}

\subsection{Transport relaxation times}
\label{ssec:Transport-relax-time}

Here we illustrate the importance of using the correct transport relaxation times (TRTs) in mobility calculations instead of the commonly used self-energy relaxation times (SERTs)~\cite{Song2017,Zhou2016,Restrepo2014}, which correspond to inverse scattering rates. The TRTs and SERTs of $n$-type PbTe resolved by acoustic, TO and LO modes are shown in Fig.~\ref{fig:sr-vs-rt}. The TRTs of LO phonons are much larger than the corresponding SERTs. This originates from the fact that the $1/|\qv|$ singularity of the polar Fröhlich e-ph matrix element does not contribute much to TRTs, since the scattering angle factor $(1-\beta_\kv^{\kv+\qv})$ becomes zero for $\qv\rightarrow 0$. Therefore, the SERTs strongly overestimates the contribution of LO phonon scattering to electronic transport. Apart from the difference in magnitude, the TRTs for LO scattering increase slightly with electronic energy, while the SERTs are nearly constant. In contrast, the TRTs for acoustic phonons are the same as the SERTs because electron-acoustic phonon matrix elements have no \qv-dependence~\cite{Ridley1999}, see Eq.~\eqref{eq:gkk-la}. The SERTs of acoustic phonons lay exactly on the top of the TRTs  in Fig.~\ref{fig:sr-vs-rt}. For TO phonon scattering, there is a slight difference between the TRTs and SERTs since electron-TO phonon matrix elements depend on \qv\ according to Eq.~\eqref{eq:gkk-lo}, but do not diverge as those due to LO scattering. Our results thus demonstrate that it is essential to use the transport relaxation times when electronic transport is dominated by LO phonon scattering, as also shown previously in the case of GaAs~\cite{Ma2018}.

Using the transport relaxation times instead of the corresponding self-energy values does not change the qualitative conclusions about the relative strengths of scattering mechanisms in $n$-type PbTe, but it does change their magnitudes, see Fig.~\ref{fig:sr-vs-rt}. Low energy electrons scattered by LO phonons still have much shorter TRTs, compared to those scattered by acoustic and TO phonons. However, for higher energies, the TRTs due to acoustic phonons are more comparable to those of LO phonons. This indicates that acoustic phonon scattering will become more important for electronic transport for high doping concentrations and temperatures. Fig.~\ref{fig:sr-vs-rt} shows that the TO contribution to the TRTs is very weak, and that TO scattering will be ineffective in a wide range of temperatures and carrier concentrations.   

\subsection{Mobility}

\begin{figure}
    \centering
    \includegraphics[width=\columnwidth]{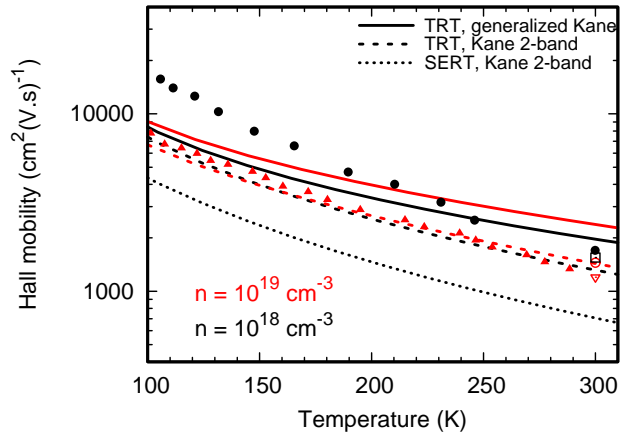}
    \caption{Temperature dependence of the Hall mobility of $n$-type PbTe calculated using the transport relaxation time (TRT) approximation with the generalized Kane model (solid lines) and with the Kane 2-band model (dashed lines). Black and red lines correspond to the doping concentrations of $10^{18}$~cm$^{-3}$ and $10^{19}$~cm$^{-3}$, respectively. Black circles and red triangles represent the experimental data from Ref.~\onlinecite{Allgaier1958} for the doping concentrations of $10^{18}$~cm$^{-3}$ and $10^{19}$~cm$^{-3}$, respectively. Dotted black line shows the mobility calculated using the self-energy relaxation time (SERT) approximation and the 2-band Kane model for the carrier concentration of $10^{18}$~cm$^{-3}$. The additional experimental data at 300~K are shown in an open square and triangle (Ref.~\onlinecite{Vineis2008}), and a circle (Ref.~\onlinecite{Harman1996}). }
    \label{fig:mobility-sr-vs-rt}
\end{figure}

We calculate the mobility of $n$-doped PbTe with doping concentrations of $\bar{n}=10^{18}$~cm$^{-3}$ and $10^{19}$~cm$^{-3}$ in the temperature range from 100~K to 300~K, \footnote{Even though PbTe is a narrow gap semiconductor, we verified that for the temperature range considered in this study, the variation of carrier densities with temperature is negligible and the electron density is very close to the doping concentration. This small magnitude of variation is expected considering the positive variation of the gap with temperature.} and compare our results with the experimental data on single-crystalline samples from Allgaier and Scanlon~\cite{Allgaier1958}. (The mobility values of Silverman and Levinstein~\cite{Silverman1954} for the $\bar{n}=10^{18}$~cm$^{-3}$ are very similar to those of Allgaier and Scanlon~\cite{Allgaier1958}).  Full and dashed red lines in Fig.~\ref{fig:mobility-sr-vs-rt} show our TRT calculations using generalized and 2-band Kane model, respectively. We note that the experimental data represent Hall mobility. We thus multiply our drift mobility with the Hall factor given as~\cite{Herring1956,Ravich1970a} 
\begin{equation}
    \frac{\sigma_H}{\sigma} = \frac{\ev{\tau^2} \times \ev{1} }{\ev{\tau}^2}  \times \frac{3K(K+2)}{(2K+1)^2},
\end{equation}
where $K = \mpara / \mperp$, and the angular brackets represent the thermal average of a function, for example $\tau$, given as $\ev{\tau} = \int d\kv   (-\partial f^0 / \partial E) |\kv|^2 \tau$.  Our estimated ratio between the Hall and drift mobility of $n$-type PbTe is $\approx 0.9$ for the doping concentrations and temperatures of interest. 

Our calculated mobility with the Kane 2-band model for the carrier concentration of $10^{19}$~cm$^{-3}$ is in very good agreement with experiment, see Fig.~\ref{fig:mobility-sr-vs-rt}. At room temperature, the calculated mobility is 1569~cm$^2$/Vs and  experimental  values are in the range from 1335~cm$^2$/Vs to 1550~cm$^2$/Vs.
The computed mobility for $\bar{n}=10^{18}$~cm$^{-3}$ and the Kane 2-band model is very similar to that for $\bar{n}=10^{19}$~cm$^{-3}$, as observed in other measurements~\cite{RAVIC1971,Harman1996,Vineis2008}. 
However, our calculation for $\bar{n}=10^{18}$~cm$^{-3}$ exhibits a departure from the measured data of Ref.~\onlinecite{Allgaier1958}, particularly for lower temperatures around $100$~K. The mobility computed using the generalized Kane model is closer to experiment for $\bar{n}=10^{18}$~cm$^{-3}$ and $T\sim 200$~K than that of the Kane 2-band model due to a lower density of states, but it is further away from experiment for $\bar{n}=10^{19}$~cm$^{-3}$. The variations of our calculated results indicate that the non-parabolicity of the conduction band L valleys has a non-negligible effect on the mobility of PbTe, which is challenging to describe accurately from first principles and translate into simpler models.  

To further understand the difference between the computed and experimental mobility values, we note that if we fit the Kane models only over a very small energy range close to the conduction band minimum, the model gives a better estimation of the experimental mobility at low $T$ for $\bar{n}=10^{18}$~cm$^{-3}$, due to smaller effective masses. However, we fit the models in a wider energy window to cover the wide range of doping concentrations up to $10^{20}$~cm$^{-3}$ in the experiments. This results in a worse fitting close to the conduction band minimum and is partially responsible for the difference in mobility between our model and the experiment at low $T$ and low doping concentration. Experimentally,
it is difficult to precisely measure the carrier concentration of lightly doped samples with $\bar{n}\sim 10^{18}$~cm$^{-3}$. Furthermore, Ref.~\onlinecite{Allgaier1958} reports that the carrier concentration decreases at lower temperatures for the sample with $\bar{n}=10^{18}$~cm$^{-3}$ carriers, and is $10\--20$\% lower at 4.2~K than at 295~K. This could partially explain a steeper mobility increase with decreasing temperature for the $\bar{n}=10^{18}$~cm$^{-3}$ sample, and its larger departure from the mobility of the sample with $\bar{n}=10^{19}$~cm$^{-3}$. We also find that the temperature dependence and the mobility values for the sample with $\bar{n}=10^{19}$~cm$^{-3}$ are consistent with the measurements for $\bar{n}=2\times 10^{19}$~cm$^{-3}$ and temperatures above $300$~K ~\cite{Pei2014}.  
We thus speculate that the experimental mobility values for the sample with $\bar{n}=10^{19}$ cm$^{-3}$ may be more accurate than those for $\bar{n}=10^{18}$ cm$^{-3}$. 
In the light of all the described theoretical and experimental challenges, we deem the agreement between our calculated and experimental mobility values reasonably good.    

We now demonstrate that using the self-energy relaxation time approximation instead of the correct transport one leads to a large underestimation of our calculated mobility values. Fig.~\ref{fig:mobility-sr-vs-rt} shows that the mobility of $n$-type PbTe obtained using the SERTs and the 2-band Kane model for $\bar{n}=10^{18}$~cm$^{-3}$ (dotted black line) is significantly lower than that computed using the TRTs (dashed  black line). This difference is $\sim$50\% in the entire temperature range considered, which is significantly larger than the sensitivity of our $\mu$ values to the non-parabolicity effects ($\sim 25$\%). Furthermore, the mobility computed using the SERTs significantly underestimates the experimental values. This result confirms the importance of using the TRTs instead of the SERTs to calculate the electronic mobility of polar materials like PbTe, where LO phonon scattering is very strong. 

\begin{figure}
    \centering
    \includegraphics[width=\columnwidth]{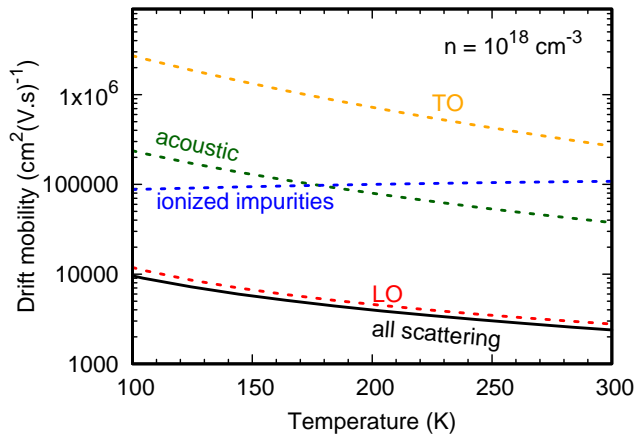}
    \caption{Drift mobility of $n$-type PbTe calculated using the generalized Kane model for the carrier concentration of $10^{18}$~cm$^{-3}$ as a function of temperature (solid black line) and its contributions due to scattering with different phonon modes and ionized impurities (colored dashed lines).}
    \label{fig:mobility-T}
\end{figure}

To identify the contribution of each scattering channel to electronic transport in $n$-type PbTe, we plot the drift mobility resolved by phonon modes in Fig.~\ref{fig:mobility-T} for $\bar{n}=10^{18}$ cm$^{-3}$. We confirm again that LO phonon scattering is the   dominant scattering channel, which is in contrast to the previous works claiming that the strongest scattering mechanism is that due to acoustic phonons~\cite{Lalonde2011,Pei2014}. In these works, the temperature dependence of the mobility suggested that charge carriers are predominantly scattered by acoustic phonons in PbTe. However, after accounting for non-parabolicity, anisotropy, exact energy conservation and the Fermi-Dirac distribution in our BTE model, we find that LO scattering has a similar temperature dependence as acoustic scattering, see Fig.~\ref{fig:mobility-T}. Acoustic and TO phonon scattering contribute little to the mobility for the carrier concentration of $10^{18}$ cm$^{-3}$ and the temperature range between 100 K and 300 K. Ionized-impurity scattering also has a small impact on the mobility due to the very large static dielectric constant of PbTe ($\epsilon_s=313.65$).

\begin{figure}
    \centering
    \includegraphics[width=\columnwidth]{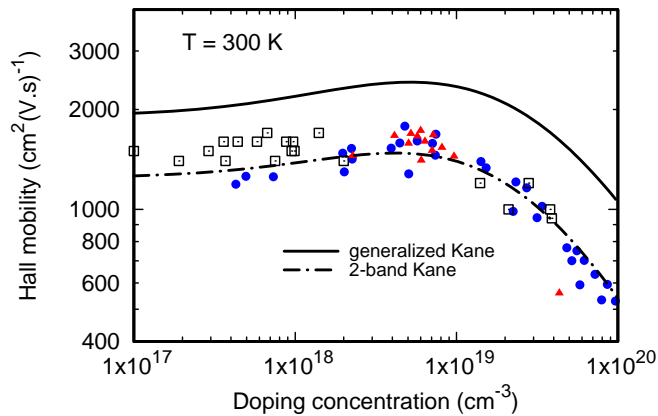}
    \caption{Hall mobility of $n$-type PbTe limited by scattering due to phonons and ionized impurities versus doping concentration calculated using the generalized and 2-band Kane model (solid and dash-dotted black lines, respectively). Experimental data are shown in circles (Ref.~\onlinecite{RAVIC1971}), squares (Ref.~\onlinecite{Harman1996}), and triangles (Ref.~\onlinecite{Vineis2008}).  }
    \label{fig:mobility-n}
\end{figure}

\begin{figure}
    \centering
    \includegraphics[width=\columnwidth]{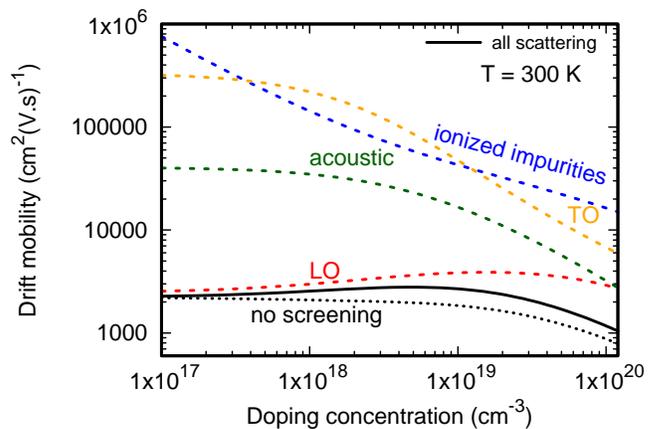}
    \caption{Drift mobility of $n$-type PbTe calculated using the generalized Kane model including screening at 300~K as a function of doping concentration (solid black line), and its  contributions due to  scattering with different phonon modes and ionized impurities (dashed colored lines). Dotted black line shows the drift mobility without screening effect.}
    \label{fig:mobility-n-bymodes}
\end{figure}

The calculated Hall mobility of $n$-type PbTe at room temperature as a function of doping concentration is illustrated in Fig.~\ref{fig:mobility-n}, and compared with experiments~\cite{RAVIC1971,Harman1996,Vineis2008}. The mobility values computed using both the generalized and 2-band Kane models are reasonably close to the measured data (solid and dash-dotted black lines, respectively). 
Fig.~\ref{fig:mobility-n-bymodes} shows the contributions to the drift mobility from each scattering channel. LO scattering is the strongest for low carrier concentrations, where the mobility is nearly insensitive to concentration variations. For higher doping concentrations, charge carriers screen the potential generated by phonons, and the e-ph scattering due to the long-range LO phonon interaction is reduced. 
Neglecting screening (dotted black line in Fig.~\ref{fig:mobility-n}) results in lower values of electronic mobility for higher doping levels with respect to the case when screening is accounted for (solid black line). However, the screening effect is not very large owing to the large high-frequency dielectric constant of PbTe ($\epsilon_\infty=34.85$). As a result, acoustic scattering becomes comparable to that of LO phonons only for very high carrier concentrations ($>$10$^{20}$~cm$^{-3}$) that screen LO phonons significantly. The effect of ionized-impurity scattering on the mobility remains relatively weak in the whole range of concentrations, and it weakens as the carrier concentration increases due to screening. In the concentration range between 10$^{19}$~cm$^{-3}$ and 10$^{20}$~cm$^{-3}$ where scattering due to acoustic modes becomes stronger, the mobility decreases rapidly with increasing carrier concentration.  

Scattering due to soft TO phonons is the weakest and almost negligible in the whole range of temperatures and carrier concentrations considered here, see Figs.~\ref{fig:mobility-T} and \ref{fig:mobility-n-bymodes}. As our symmetry analysis showed, the weak electron-TO phonon scattering is caused by the symmetry-forbidden scattering between the conduction band states exactly at the L point and the TO modes at the zone center. The low-lying soft TO modes are beneficial for the thermoelectric performance of PbTe since they cause strong acoustic-TO scattering and reduce the lattice thermal conductivity, while not affecting significantly the electronic transport properties. Our findings suggest that other materials with soft phonon modes and symmetry-forbidden electron-soft phonon coupling may also have good thermoelectric properties.

%%%%%%%%%%%%%%%%%%%%%%%%%%%%%%%%%%%%%%%%%%
%%   Conclusion
%%%%%%%%%%%%%%%%%%%%%%%%%%%%%%%%%%%%%%%%%%
\section{Conclusion}

We develop an electronic transport model for $n$-type PbTe based on the Boltzmann transport equation in the transport relaxation time approximation, where all the parameters are obtained from {\it ab-initio} calculations that accurately represent the dispersion  of the electronic bands near the band edge. We obtain a very good agreement between our computed electronic mobility with experimental data in a range of temperatures and carrier concentrations. We find that charge carriers are predominantly scattered by longitudinal optical phonons in $n$-type PbTe. Acoustic phonon scattering becomes important only for very high concentrations ($\sim$10$^{20}$~cm$^{-3}$) when LO phonons are sufficiently screened. Scattering due to soft transverse optical phonons is weak owing to the symmetry-forbidden scattering between the conduction band states at L and the zone center soft modes. This finding suggests that soft optical modes are the main reason for the excellent thermoelectric performance of $n$-type PbTe: they cause the low lattice thermal conductivity while not degrading the electronic conductivity. Our results indicate that other high-performing thermoelectric materials may be found among materials with soft modes that are weakly coupled to conducting electronic states.

\section*{Acknowledgement}
We would like to thank Nenad Vukmirovi\'c, Felipe Murphy-Armando and G. Jeffrey Snyder for useful discussions. This work was supported by Science Foundation Ireland (SFI) under Investigators Programme No. 15/IA/3160. We acknowledge the Irish Centre for High-End Computing (ICHEC) for the provision of computational facilities.

\appendix

%------------------------------------------------------------
%  Kane model
%------------------------------------------------------------
\section{Kane model}
\label{apx:kane}

Using the Kane model~\cite{Kane1957,RAVIC1971}, the energy dispersion of the bottom conduction band of PbTe near the L point in the [111] direction is given as 
\begin{equation}
\frac{\hbar^2}{2} \left( \frac{\kpara^2}{\mpara} + \frac{\kperp^2}{\mperp} \right)
=  E \left( 1 + \alpha E \right)\equiv \gamma(E),
\end{equation}
where $\alpha$ is the non-parabolicity parameter, \kpara\ and \kperp\ are the components of the wavevector parallel and perpendicular to the [111] direction, and \mpara\ and \mperp\ are the parallel and perpendicular effective masses. In our transport calculations, we also consider the contribution of the top valence band near the L point, whose dispersion is accounted for in the Kane model as a mirror image of the conduction band with respect to the middle of the direct band gap, whose energy is given by $E_h = -(E + \Eg)$. Furthermore, we use the same parameters to describe e-ph scattering for the valence band as those calculated for the conduction band, given in Table~\ref{tab:parameters}. These approximations in the valence band description are reasonable since we find that the valence band has a negligible effect on the mobility values reported in the paper. 

To perform all the summations given by Eqs.~\eqref{eq:cond}, \eqref{eq:TRTA} and \eqref{eq:rate} using the Kane band structure, we first define a new coordinate system by scaling the $\kv$ components along the three principal axes of the ellipsoids so that the energy surfaces are spherical:
\begin{equation}
	w_i = k_i / {m^*_i}^{1/2} ,
\end{equation}
where $i$ denotes the direction. Consequently,
\begin{equation}
	\hbar^2 w^2 / 2 = E (1 + \alpha E).
\end{equation}
The conduction band density of states (DOS) at electron energy $E$ is then given as
\begin{equation}
    D(E) = \frac{m_d^{3/2}}{\sqrt{2} \pi^2 \hbar^3}   \sqrt{\gamma(E)} \frac{d \gamma(E)}{dE},
\end{equation}
where $m_d=({\mperp}^2\mpara)^{1/3}$ is the DOS effective mass.

The gradient of energy with respect to the wave vector, which is proportional to the group velocity, reads as
\begin{equation}
    \grad_\kv E = \frac{d E}{d \gamma} \hbar^2 \left( \frac{k_x}{\mperp},\frac{k_y}{\mperp},\frac{k_z}{\mpara} \right),
\end{equation}
in the coordinate system where the $z$-axis is parallel to the $[111]$ direction. From here, the scattering angle factor $\beta$ determining transport relaxation times can be written as 
\begin{equation}
 \label{eq:betakk}
\beta_w^{w'} = \frac{ \cos\theta + a \cos\delta \cos\delta'}{\sqrt{1 + a \cos^2\delta}\sqrt{1 + a \cos^2\delta'}},
\end{equation}
where $a=(\mperp/\mpara)^2-1$, $\theta$ is the angle between $\kv$ and $\kv'$, and $\delta$ ($\delta'$) is the angle between $\kv$ ($\kv'$) and the parallel axis of the ellipsoidal isoenergy surface.

We also derive the overlap integral, which enters the expressions for electron-phonon matrix elements, from the \kdotp\ Hamiltonian as
\begin{equation}
\label{eq:Ikk}
I_{w,w'} = a_w a_{w'} +\frac{ c_w c_{w'}  (\cos \theta + b \cos \delta \cos \delta' )}{ \sqrt{1+ b \cos^2 \delta} \sqrt{1+ b \cos^2 \delta'}  } ,
\end{equation}
where $b = \mperp / \mpara -1$, the angles $\theta$, $\delta$, and $\delta'$ are defined in the same way as in Eq.~\eqref{eq:betakk}, and the factors $a_w$ and $c_w$ are given by
\begin{equation}
\begin{cases}
a_w  &=\left( \frac{1 + \alpha E}{1 + 2 \alpha E} \right)^{1/2} , \\
c_w  &=\left( \frac{\alpha E}{1 + 2 \alpha E} \right)^{1/2}     . \\
\end{cases}
\end{equation}

In the expressions for calculating e-ph matrix elements, we also need to re-write the phonon wave vector $\qv = \kv'-\kv$ in the scaled $w$ coordinate system. Starting from
\begin{equation}
	q^2 = \norm{\kv'-\kv}^2 = \norm{\kv'_\perp-\kv_\perp}^2 + (k'_\parallel-k_\parallel)^2 ,
\end{equation}
we replace the parallel and perpendicular components and write
\begin{equation}
\label{eq:q2}
\begin{split}
    q^2 =& \mperp  (w'^2 + w^2 - 2 w w' \cos \theta)   \\
		 &		+ (\mpara - \mperp) \times (w' \cos \delta' - w \cos \delta)^2.
\end{split}
\end{equation}

The transport relaxation time given by Eq.~\eqref{eq:TRTA} can be rewritten in the $w$ coordinate system as~\cite{Olechna1962}:
\begin{multline}
\tau_{w, \lambda}^{-1} = \frac{1}{(2\pi)^3}  \int_0^{2\pi} d\phi' \int_{-1}^{1} d(\cos \theta') [1 - \ev{\beta(\theta')}] \\
\times \biggl[\frac{ w^2(E'_{-}) }{[\partial E / \partial w]_{w=w(E'_{-}) } }  \frac{1-f^0(E'_{-})}{1-f^0(E)}  \ev{{S_{\lambda}}}(w(E'_{-}) ,w,\theta') \\
+\frac{ w^2(E'_{+}) }{[\partial E / \partial w]_{w=w(E'_{+}) } }  \frac{1-f^0(E'_{+})}{1-f^0(E)}  \ev{{S_{\lambda}}}(w(E'_{+}) ,w,\theta')  \biggr],
\end{multline}
where $\lambda$ denotes scattering due to different phonon modes or impurities, $\theta'$ is defined as the angle between the initial and final $w$ vectors, and $E'_{+/-}$ is the electron energy of the final state due to phonon absorption/emission. For LO and TO phonons, the energies $E'_{+/-}$ are obtained from the exact solution of the energy conservation $E(w'_\pm) = E(w) \pm \hbar\omega_{\qv\lambda}$, where $q$ is given by Eq.~\eqref{eq:q2}. The analytical forms for electronic and phonon bands enable us to solve the energy conservation exactly, using the linear search and bisection methods. For acoustic phonons and impurities, the energies of the initial and final states are the same. $S_\lambda$ represents the scattering term given by Eq.~\eqref{eq:rate} without its energy-conserving $\delta$-functions that are calculated by integrating Eq.~\eqref{eq:rate} over the radial coordinate in $w$ space and inserting the density of states factor $[\partial E / \partial w]^{-1}$. In general, $S_\lambda$ depends on $w$, $w'$, $\theta'$ and the angles  $\delta$ and $\delta'$ between $w$ and $w'$ vectors and the parallel axis, respectively. Here we calculate its angular average value $\ev{{S_\lambda}}(w',w, \theta')$ over $\delta$ and $\delta'$, which ``folds" the anisotropic effects from $\delta$ and $\delta'$ into an ``effective'' isotropic average~\cite{Olechna1962}. To calculate $\ev{S}(w',w, \theta')$, we replace the overlap integral $|I|^2$ and $q^2$ in the matrix element expressions (given by Eqs.~\eqref{eq:gkk-la}, \eqref{eq:gkk-lo},  \eqref{eq:gkk-froh} and \eqref{eq:gkk-ii}) by their angular averages over $\delta$ and $\delta'$. In the same spirit, the angular average of $q$ is used to determine the phonon frequency in the energy conservation condition. The scattering angle factor $\beta$ is also replaced by its angular average. These averages are evaluated numerically using Eqs.~\eqref{eq:betakk}, \eqref{eq:Ikk} and \eqref{eq:q2}. 

Finally, the electronic conductivity expression given by Eq.~\eqref{eq:cond} is expressed as 
\begin{equation}
\label{eq:cond2}
    \overline{\sigma} = \frac{ e^2 N_v^L m_d^{3/2}}{\pi^2 } \int_{0}^{\infty} \left(-\pdd{f^0}{E} \right) \tau_{w} w^2 \overline{v}^2 dw ,
\end{equation}
where the $N_v^L=4$ is the number of L valleys, $f^0$ the Fermi-Dirac distribution, $\overline{v}^2 = \sum_i v_i^2 /3$ the average group velocity, and $\tau_w^{-1} = \sum_\lambda \tau_{w,\lambda}^{-1}$. The conductivity per phonon mode is obtained using Eq.~\eqref{eq:cond2} by replacing $\tau_w$ with $\tau_{w,\lambda}$. The hole conductivity is obtained by replacing $E$ in Eq.~\eqref{eq:cond2} with the valence band dispersion $E_h=-(E+\Eg)$. The total conductivity is the sum of the electronic and hole conductivities.
Using the averaging procedure, the numerical evaluation of Eq.~\eqref{eq:cond} is simplified from a five-dimensional integration to a two-dimensional integration (an ``outer'' loop over $w$ for the conductivity and an ``inner'' loop over $\theta'$ for the transport relaxation times), and contains additional two-dimensional integrals for the calculation of the isotropic averages. In all our calculations, we have used 50 points on a Gaussian quadrature grid for the angles $\theta$, $\phi$, $\delta$ and $\delta'$, and 1000 equally spaced points for $w$ in the energy range of $(\pm10 k_B T)$ around the Fermi level. 

%------------------------------------------------------------
%  Ionized Impurity
%------------------------------------------------------------
\section{Ionized impurity scattering}
\label{apx:ionized-impurity}

We consider that chemical doping introduces charged impurity scattering in addition to changing the Fermi level in a material. We treat the ionized impurity scattering as elastic and use the Brooks-Herring model~\cite{Ridley1999a,Sofo1994}. In this approach, the matrix-element for $N_i$ ionized impurities of charge $Z_{\rm II} e$ is written as
\begin{equation}
  \label{eq:gkk-ii}
|g_{\rm II}|^2 = \frac{e^4 N_i Z^2_{\rm II}}{ V \epsilon^2_s} \frac{1}{\left[q^2 + (q_{\rm scr}^{\rm s})^2\right]^2}  I^2_{\kv,\kv+\qv} ,
\end{equation}
where $V$ is the unit cell volume, $\epsilon_s$ is the static dielectric constant, and  $q_{\rm scr}^{\rm s}$ is the screening wave vector. $Z_{\rm II}$ is taken to be 1, and $N_i$ is chosen to be equal to the electron concentration, thus assuming full ionization~\cite{RAVIC1971}.

%------------------------------------------------------------
% Charge screening 
%------------------------------------------------------------
\section{Charge screening}
\label{apx:screening}
In the Fr{\"o}hlich and Brook-Herring models of scattering due to polar optical phonons and ionized impurities, respectively, the presence of conduction band electrons screens the long-range electric field via the screening wave vector $q_{\rm scr}^{\infty}$ and $q_{\rm scr}^{\rm s}$, respectively. According to the work of Hauber and Fahy \cite{Hauber2017} using coupled plasmon-phonon collective modes, the effect of screening in PbTe is fairly weak for electron density up to $10^{20}$~cm$^{-3}$. In addition, the LO phonon frequency is low in PbTe. Therefore, we computed the screening vectors using Thomas-Fermi theory~\cite{Ridley1999b}: 
\begin{equation}
\label{eq:qscrn}
	(q^{\infty/s}_{\rm scr})^2 = \frac{-e^2}{\epsilon_0 \epsilon_{\infty/s}}  \pdd{n}{E_f} = \frac{e^2}{\epsilon_0 \epsilon_{\infty/s}} \int_0^\infty  \left( - \pdd{f^0}{E}  \right) D(E) dE.
\end{equation}
For LO scattering, we do not take into account screening due to polar phonons and use the high-frequency dielectric constant $\epsilon_{\infty}$ in the expression for the screening wave vector $q_{\rm scr}^{\infty}$. In contrast, screening due to polar phonons is included when calculating ionized impurity scattering by using the static dielectric constant $\epsilon_s$ in the expression for the screening wave vector $q_{\rm scr}^{\rm s}$.

Screening also induces a \qv\ dependence of the LO phonon frequencies near $\Gamma$. In the limit of very high doping and complete screening, the LO frequency approaches that of TO mode for $\qv \rightarrow 0$, and the unscreened LO frequency for $q \gg q_{\rm scr}$. In this case, the \qv\ dependence of the LO phonon frequency can be described with a quadratic function for $\qv \rightarrow 0$ ($\omega^{\rm LO}_\qv=\omega^{\rm LO}_{\Gamma}+\partial^2\omega^{\rm LO}/\partial |\qv|^2 \times |\qv|^2$) and fitted to the DFPT-LDA phonon bands obtained by setting the Born effective charge to zero (the fit was done along $\Gamma-L$, $\Gamma-X$ and $\Gamma-K$ directions). We estimate that the impact of including this \qv\ dependence on the scattering rate for the carrier concentration of $10^{20}$~cm$^{-3}$ is noticeable only very close to the band edge, and makes a difference of less than 2\% to the mobility. Therefore, we neglect the \qv\ dependence of LO frequencies due to screening, and approximate it to be equal to the unscreened value of the LO frequency at the zone center.

%------------------------------------------------------------
% El-ph matrix elements from DFPT 
%------------------------------------------------------------
\section{Electron-phonon matrix elements from density functional perturbation theory}
\label{apx:el-ph}

Within density functional perturbation theory (DFPT), the electron-phonon matrix element for an electron scattering event from a state \kv\ and band $n$ to a state \kv+\qv\ and band $m$ via a phonon with wavevector $\qv$ and branch $\lambda$ can be defined as \cite{Giustino2017}:
\begin{eqnarray}
\label{ch2_elphmatelement}
H_{mn}(\kv; \qv\lambda) = \left(\frac{\hbar}{2\omega_{\qv\lambda}}\right)^{1/2} \sum_{b,i}\left( \frac{1}{m_{b}} \right)^{\frac{1}{2}} e^{\qv\lambda}_{b,i}\nonumber\\ \times  \left<u_{\two}\right| \partial_{b,i,\qv}v^{\text{KS}}\left|u_{\one}\right>_{\text{uc}},
\end{eqnarray}
where $e^{\qv\lambda}_{b,i}$ is the $i$-th Cartesian component of the phonon eigenvector for an atom $b$ with mass $m_b$. The subscript ``uc'' in Eq.~\eqref{ch2_elphmatelement} indicates that the integral is carried out within one unit cell. $u_{\one}$ is normalized to unity in the unit cell, and is the lattice periodic part of the wavefunction $\psi_{\one}$ expressed in Bloch form as $N_l^{-1/2}u_{\one}e^{i\kv\cdot{\bf r}}$, where $N_l$ is the number of primitive cells. $\partial_{b,i,\qv}v^{\text{KS}}$ is the lattice periodic part of the perturbed Kohn-Sham potential expanded to first order in the atomic displacement, see Ref.~\onlinecite{Giustino2017} for further details.

\bibliography{biblio}

\end{document}